\documentstyle[12pt]{article} 
\begin{document} \begin{titlepage}
\begin{flushright} UTPT-94-25 \\ hep-ph/9410324 \\ October 1994
\end{flushright} \vspace{24pt} \begin{center} {\LARGE What can a relativistic
quark model\\ tell us about charmed mesons?} \\ \vspace{40pt} {\large M.
Sutherland$^{\rm a}$, B. Holdom$^{\rm b}$, S. Jaimungal$^{\rm c}$ and Randy
Lewis$^{\rm d}$ } \vspace{0.5cm}

{\small Department of Physics, University of Toronto\\ 60 St. George St.,
Toronto, Ontario\\Canada M5S 1A7}

\vspace{12pt}

\begin{abstract}   A relativistic quark model is extended to incorporate
chiral and gauge symmetries.  We obtain the $DD^{\ast}\pi$ and
$DD^{\ast}\gamma$ couplings and find that the ratio
$\Gamma(D^{\ast 0}\rightarrow D^0 \pi^0)/\Gamma(D^{\ast 0}\rightarrow D^0
\gamma)$ constrains the charm-quark mass close to 1.45 GeV.  Large $1/m_c$
corrections appear in the heavy-quark contribution to the $DD^{\ast}\gamma$
coupling.  The model is extended further to describe seven excited
$D$ meson states.  We find that semileptonic $B$ decays into the ground and
excited
$D$ meson states do not account for the total semileptonic decay width of the
$B^0$.  The nonresonant contributions to the processes
$\overline{B}^0\rightarrow D^{(\ast)}\pi\ell
\overline{\nu}$ appear to be large enough to account for the discrepancy.

\end{abstract} \end{center}

\vspace{24ex}  {\small $^{\rm a}$ marks@medb.physics.utoronto.ca

$^{\rm b}$ holdom@utcc.utoronto.ca

$^{\rm c}$ jaimung@physics.ubc.ca

$^{\rm d}$ lewis@oldkat.physics.utoronto.ca} \end{titlepage}

\newpage \section*{Introduction}

We have recently developed a fully relativistic model of hadrons containing
one heavy quark \cite{simply}.  Hadronic matrix elements are represented by
quark loop graphs with momentum dependent interaction vertices between
hadron, heavy quark and light quark.  The only parameters in the model are
the quark masses.   We previously applied the model to
$\overline{B}\rightarrow D^{(\ast)}\ell \overline{\nu}$
\cite{hyperfine,linear},
$\overline{B}\rightarrow K^{\ast}\gamma$ \cite{kstargamma}, $\Lambda_b
\rightarrow \Lambda_c \ell \overline{\nu}$ \cite{lambdabaryons}, and
$\Omega_b^{(\ast)} \rightarrow \Omega_c^{(\ast)} \ell \overline{\nu}$
\cite{omegabaryons}.   The model does not rely on any expansion in inverse
powers of heavy-quark masses $m_Q$. However, QCD imposes severe constraints
on the form of the heavy-quark expansion, and not all models automatically
satisfy these constraints \cite{neubert}.  We have verified that the
heavy-quark expansion of our model does satisfy these constraints.

Confinement is modeled in a simple way; we simply drop the imaginary parts of
our loop graphs.  This prescription preserves the symmetry structure of the
theory, and our hypothesis is that it may give reasonable results as long as
external momenta of a diagram are such as to be far removed from whatever
unphysical singularites the diagram possesses.  (A discussion of the
singularity structure of our three-point functions, including anomalous
thresholds, may be found in \cite{anom}.)  We are finding that this simple,
most naive model of confinement is yielding some very encouraging results.
It also acts as a useful preview to the general Bethe-Salpeter approach.

Although our results are based only on a model, they suggest that some
caution is required in the use of more systematic attempts to extract
information from QCD.  In particular, some important results in the literature
depend on an expansion in powers of $1/m_c$.   We question whether this
expansion is sufficiently understood and under control at present.  Our model
provides an explicit example of how far from the heavy quark limit the charm
mass could, in fact, be.  We will later discuss an example of a quantity
receiving large $1/m_c$ corrections, and which also contributes to a measured
branching ratio.

One of the prime applications of the heavy quark expansion is in the
extraction of $|V_{cb}|$ from the processes $\overline{B}\rightarrow
D^{(\ast)}\ell \overline{\nu}$ and $\Lambda_b \rightarrow \Lambda_c \ell
\overline{\nu}$.  In the heavy-quark limit $m_{b,c} \rightarrow \infty$,
these processes are normalized in a model-independent way at zero hadronic
recoil.  But for any realistic $m_c$ in our model there are always large
positive deviations in at least one of $\overline{B}\rightarrow D^{\ast}\ell
\overline{\nu}$ or $\overline{B}\rightarrow D\ell \overline{\nu}$
\cite{hyperfine}, concurrently with rather small deviations in $\Lambda_b
\rightarrow \Lambda_c \ell \overline{\nu}$ \cite{lambdabaryons}.  The large
deviations are a sensitive function of the quark masses, and their origin
lies in the hyperfine interaction which splits the ground-state pseudoscalar
and vector $D$ mesons.  This would lead to a discrepancy between the values
of $|V_{cb}|$ extracted from these various processes under the assumption that
the deviations from the heavy-quark limit are negligible.

In Fig. 1 we show that our model's prediction for the {\em shape} of the
decay spectrum of $\overline{B}\rightarrow D^{\ast}\ell\overline{\nu}$ agrees
well with the most recent CLEO data \cite{cleo}.  If we ignored our
correction to the overall normalization and forced the heavy-quark limit
normalization at zero recoil we would obtain $|V_{cb}|=0.036$.  Our
correction to the normalization implies that the actual $|V_{cb}|$ could be
10\% or 15\% smaller.

This paper is divided into three parts.  In Parts I and II we describe two
major extensions of the model.   In Part I we incorporate chiral and gauge
symmetries to deal with various processes involving pions and photons.   In
particular the ratio $\Gamma(D^{\ast 0}\rightarrow D^0 \pi^0)/\Gamma(D^{\ast
0}\rightarrow D^0 \gamma)$ is well measured experimentally and provides a
sensitive probe to the physics of the light degrees of freedom in the $D$
mesons.  This will help constrain the only parameters appearing in our model,
the quark masses.  Our model reproduces this branching ratio for a choice of
quark masses very close to what we have previously used, and in particular it
contrains our model $m_c$ to be close to 1.45 GeV.  We have observed before
that $m_b-m_c$ was much more constrained by our model than
$m_b+m_c$; the constraint reads $m_b-m_c=3.36\pm 0.03$ GeV (see the
wedge-shaped region in Fig. 6 of \cite{linear}). With our new constraint on
$m_c$, our model $m_b$ must be close to our previously-used value of 4.8 GeV.

One ingredient in our calculation of $\Gamma(D^{\ast 0}\rightarrow D^0
\pi^0)/\Gamma(D^{\ast 0}\rightarrow D^0 \gamma)$ is the heavy-quark
contribution to the $D^\ast D\gamma$ coupling. We find large $1/m_c$
corrections to this quantity, which are analogous to the large corrections we
find in semileptonic $B$ decays.  For the $D^\ast D\pi$ and $D^\ast D\gamma$
couplings we find values which are one-third as large as the nonrelativistic
quark model predictions.  The small $D^\ast D\pi$ coupling has implications
for calculations in heavy-quark chiral perturbation theory, and in particular
the quantity
$\Delta_D\equiv (M_{D_s^{\ast}} - M_{D_s})- (M_{D^{\ast}} - M_{D})$.  We also
include a discussion of charge radii.

We then consider the nonresonant contributions to the processes
$\overline{B}^0\rightarrow D^{(\ast)}\pi\ell \overline{\nu}$ in the context of
the problem of the missing exclusive semileptonic $\overline{B}^0$ decay
modes.   The branching fraction for $\overline{B}^0\rightarrow
\ell\overline{\nu}+{\rm anything}$ is
$(9.5\pm1.6)\%$, while those for the dominant exclusive modes
$\overline{B}^0\rightarrow D^+\ell\overline{\nu}$ and
$\overline{B}^0\rightarrow D^\ast(2010)^+\ell\overline{\nu}$ are
$(1.9\pm0.5)\%$ and
$(4.4\pm0.4)\%$, respectively \cite{pdg}.  The model indicates that the
nonresonant contributions play an important role in accounting for the
discrepancy.

In Part II we extend the model to include seven excited charmed mesons.  From
the masses (or estimates of the masses) of these mesons we are able to
determine appropriate model vertex functions.   We calculate and display the
various amplitudes for semileptonic $B$ meson decays into these excited
states.  Our results for the various decay widths and electron energy spectra
are in qualitative agreement with those of other authors.  We confirm in
particular that the decay rates to excited $D$'s are too small to account for
the above mentioned discrepancy in semileptonic $B$ decays.

In Part III we present some details of the heavy-quark limit of the model as
it applies to the excited $D$'s.  The full-model form factors deviate
considerably from their values in the heavy-quark limit, as expected since
the masses of these states are significantly greater than the charm-quark
mass.  We include the heavy-quark-limit results primarily for comparison with
other models.

Also in Part III we study two modifications of the model.  We first study the
possible modification necessary to describe strange mesons, and in particular
the process $B^+ \rightarrow \overline{K}^{\ast 0}e^+\nu$.  We address the
question of whether the $s$ should be treated as heavy or light. And second,
we investigate the effect of modifying the form of the momentum damping at
the basic quark-antiquark-meson vertex of the model.  We consider an
exponential damping factor, and we recover our previous conclusion that there
is always a large positive deviation from the heavy-quark limit in one of
$\overline{B}\rightarrow D^{\ast}\ell \overline{\nu}$ or
$\overline{B}\rightarrow D\ell \overline{\nu}$.

\section*{Part I} \section*{Gauge and chiral symmetries}

We begin with the light-quark triplet and triplets of pseudoscalar or vector
meson fields $H_Q$ with heavy quark $Q$. \begin{equation}
\begin{array}{cc} Q & H_Q \sim (Q\overline{u},Q\overline{d},Q\overline{s}) \\
\hline b & (B^-,\overline{B}^0,\overline{B}^0_s) \\ c & (D^0,D^+,D^+_s)
\end{array}
\;\;{\rm and}\;\;  q=\left(\begin{array}{c} u\\d\\s \end{array}\right)
\end{equation}

Under local SU(3)$_{V\times A}$ transformations with external vector and
axial-vector fields $V_{\mu}$ and $A_{\mu}$ transforming as \begin{equation}
V_{\mu}+A_{\mu}\gamma_5 \rightarrow
G(V_{\mu}+A_{\mu}\gamma_5)G^{\dagger}+iG\partial_{\mu}G^{\dagger},
\end{equation} the quark fields transform according to $Q\rightarrow Q$ and
$q \rightarrow Gq $, where $ G=\exp (i[\theta_V+\theta_A \gamma_5])$.

We define $\xi=\exp(i{\cal M}\gamma_5/f_{\pi})$, where $f_{\pi}\simeq 132$
MeV and \begin{equation}  {\cal M}=\left(\begin{array}{ccc}
\frac{1}{\sqrt{2}}\pi^0+\frac{1}{\sqrt{6}}\eta & \pi^+ & K^+ \\ \pi^- &
-\frac{1}{\sqrt{2}}\pi^0+\frac{1}{\sqrt{6}}\eta & K^0 \\ K^- & \overline{K}^0
& -\frac{2}{\sqrt{6}}\eta \end{array}\right) .\end{equation}  Then $\xi$
transforms as $\xi \rightarrow g \xi G^{\dagger}$.

The gauge-invariant version of the interaction Lagrangian is
\begin{equation} {\cal L}_{\rm int} = \sum_{Q=b,c} \sum_{H} \int dy
\;\overline{Q}(x){\cal L}_H(x,y) F_{H}(x-y) + {\rm h.c.},
\label{ww}\end{equation} where the function $F_{H}(x-y)$ is the Fourier
transform of the damping factor, \begin{equation} F_{H}(x-y)=
\int \frac{d^4k}{(2\pi)^4}
e^{-ik\cdot(x-y)}\left(\frac{Z_H^2}{-k^2+\widehat{\Lambda}_H^2}\right)^n .
\label{damping} \end{equation} For the ground-state mesons, $n=1$ and ${\cal
L}_H(x,y)$ are given by \begin{equation} \begin{array}{cc} H & {\cal
L}_H(x,y)\\ \hline 1\;0^-_{1/2} & -iH_Q(x) \gamma_5 K(x,y)\xi(y) q(y) \\
1\;1^-_{1/2} & iH\!\!\!\!\!/\;_Q(x) K(x,y)\xi(y) q(y) . \end{array}
\label{LintSU3} \end{equation}

The quantity $K(x,y)$ is a path-ordered exponential \begin{equation}
K(x,y)={\rm P}\exp\left\{i\int_y^x dz^{\mu}\, \Gamma_{\mu}(z) \right\}
,\end{equation} where \begin{equation} \Gamma_{\mu}=\frac{1}{2}\xi
[i\partial_{\mu}+V_{\mu}+A_{\mu}\gamma_5]\xi^{\dagger}+\frac{1}{2}
\xi^{\dagger}[i\partial_{\mu}+V_{\mu}-A_{\mu}\gamma_5]\xi \end{equation}
Under the SU(3)$_{V\times A}$ transformations, $K(x,y) \rightarrow
g(x)K(x,y)g^{\dagger}(y)
$, and the meson fields transform as $ H_Q \rightarrow H_Q g^{\dagger}$.

Mass terms involve an SU(3)$_{V\times A}$-invariant piece and an
SU(3)$_{V\times A}$-breaking piece.
\begin{equation} {\cal L}_{\rm mass} = -\overline{q}(m\xi^2+{\cal M}_q) q ,
\end{equation} where we take $m\sim 250$ MeV and $ {\cal M}_q = {\rm
diag}(m_u^{\rm curr},m_d^{\rm curr},m_s^{\rm curr})\sim (0,0,170)$ MeV. When
${\cal M}_q=0$, the full Lagrangian \begin{equation}{\cal
L}=\overline{Q}(i\partial\!\!\!/-m_Q)Q +
\overline{q}(i\partial\!\!\!/+V\!\!\!\!/+ A\!\!\!/ \gamma_5) q +{\cal L}_{\rm
mass} + {\cal L}_{\rm int} \label{chiral} \end{equation} is invariant under
the local SU(3)$_{V\times A}$ transformations.

\section*{$D^{\ast}\rightarrow D\pi$ and $D^{\ast}\rightarrow D\gamma$}

In this section we describe our calculation of $\Gamma(D^{\ast 0}\rightarrow
D^0 \pi^0)/\Gamma(D^{\ast 0}\rightarrow D^0 \gamma)$.  This quantity depends
sensitively on the only two parameters appearing in our model for $D$ mesons,
the charm mass and the light quark mass
$m$.  We will show that a parameter choice which gives an acceptable result is
$m_c=1450$ MeV and $m=270$ MeV.  The main point is that these values are close
to the values,
$m_c=1440$ MeV and $m=250$ MeV, appearing in our previous work.  For
consistency with our previous results we will continue to use our old
parameter choice in all other sections of this paper.  Except where noted,
our results in other sections do not display such a large sensitivity to
quark masses.

In this section we shall also be comparing the $D$ meson results to $B$ meson
and heavy-quark limit results.  For the $b$ quark mass we adopt our previous
value, $m_b=4.8$ GeV, as explained above.

We first consider the matrix element \begin{equation}f_{\pi}\langle
H^0(M^{\prime}v^{\prime})\pi^+(l) | H^{\ast+}(Mv,\varepsilon) \rangle =
\sqrt{MM^{\prime}}(\omega+1)h^{H\pi}(\omega)l\cdot
\varepsilon .\end{equation}  We define $g_{H^{\ast}H\pi}\equiv h^{H\pi}(1)$.
This can be calculated using the pion couplings derived above, or it can be
extracted more simply by inserting the axial current on the light-quark
line.  Our results are
$g_{D^{\ast}D\pi}=0.28$ and $g_{B^{\ast}B\pi}=0.32$.  The first two terms in
the heavy-quark expansion are
\begin{equation} g_{H^{\ast}H\pi}=0.34-0.17
\frac{\overline{\Lambda}}{m_Q}, \label{aa}\end{equation} where
$\overline{\Lambda}\equiv M - m_Q \simeq 500$ MeV is the lowest-order
difference between the meson and heavy-quark masses.  These numbers are in
agreement with other relativistic models and sum rules
\cite{CFN}\cite{gatto}\cite{bardeen}, and they are in marked contrast to the
nonrelativistic quark model which gives a value of unity.

We next consider the matrix element of the electromagnetic current, and we
distinguish the light- and heavy-quark contributions.
\begin{equation}
\langle H_Q(M^{\prime} v^{\prime}) | J_{\mu}^{\rm
e.m.}|H_Q^{\ast}(Mv,\varepsilon)
\rangle = \sqrt{M M^{\prime}} [e_q h^{H\gamma}_q(\omega) + e_Q
h^{H\gamma}_Q(\omega) ] i\varepsilon_{\mu\nu\rho\sigma}\varepsilon^{ \nu}
v^{\prime \rho} v^{\sigma} .\end{equation} For the light-quark contribution we
find
$h^{D\gamma}_q(1)=1.56$ and
$h^{B\gamma}_q(1)=5.00$.  The first two terms in the heavy-quark expansion
are \begin{equation} h^{H\gamma}_q(1) = 0.50
\frac{m_Q}{\overline{\Lambda}} \left( 1+0.47\frac{\overline{\Lambda}}{m_Q}
\right). \end{equation}   The light-quark contribution is not predicted by
heavy-quark symmetry, but the nonrelativistic quark model predicts a value
for $h^{H\gamma}_q(1)$ of order the heavy-quark to light-quark mass ratio.
Our result is about a factor of three smaller, which mirrors the situation
with the pion coupling.  In both cases, the relevant diagram involves either
a vector or axial coupling to the light quark line.  The suppression
associated with the relativistic nature of the diagram seems to apply to both
cases equally.  We also note that in both cases the heavy-quark expansion
provides a good representation of the full-model results.

The situation is different for the heavy-quark contribution to the
$DD^{\ast}\gamma$ couplings. We find
$h^{D\gamma}_Q(1)=1.54$ and
$h^{B\gamma}_Q(1)=1.16$, to be compared with the first two terms in the
heavy-quark expansion
\begin{equation}h^{H\gamma}_Q(1)=1+\frac{\overline{\Lambda}}{m_Q}(1-\xi_3(1)).
\end{equation}    The latter gives $h^{D\gamma}_Q(1)=1.35$ and
$h^{B\gamma}_Q(1)=1.10$. ($\xi_3(1)$ is negligible in our model
\cite{simply}.)  The deviations from the heavy-quark expansion are much more
significant here than they are for the pion and photon couplings to the light
quark.  As we have said, these large corrections are analogous to the large
$1/m_c$ corrections we find for $V_{cb}$.

We now use the above results to extract the $D\pi$ and $D\gamma$ decay widths
of the $D^\ast$. We find $\Gamma(D^{\ast +}\rightarrow D^0
\pi^+) + \Gamma(D^{\ast +}\rightarrow D^+ \pi^0)= 20.9$ KeV and
$\Gamma(D^{\ast 0}\rightarrow D^0 \pi^0)= 9.4$ KeV.   For the radiative decay
branching fractions we also display the effect of the leading SU(3) breaking
corrections calculated in \cite{everybody}.  We obtain \begin{equation}
\begin{array}{cccc} &{\rm experiment}&{\rm with\,corrections } &{\rm without\,
corrections}\\ {\rm B}\left({{D}^{\ast 0}\rightarrow{D}^{0}\gamma
}\right)(\%)&36.4\pm 2.8&37.8&45.6\\ {\rm B}\left({{D}^{\ast
+}\rightarrow{D}^{+}\gamma }\right)(\%)&1.1\begin{array}{c}+1.4\\
-0.7\end{array} &2.9&2.1\end{array}\label{bb} \end{equation}  The agreement
in the first two columns is a reflection of the quark masses we have chosen.

We consider briefly the implications for the $SU(3)$-violating quantities
$\Delta_H\equiv (M_{H_s^{\ast}} - M_{H_s})- (M_{H^{\ast}} - M_{H}).$   It was
stressed in \cite{randall} that the measured values,
$\Delta_D=0.9\pm 1.9$ MeV and $\Delta_B=1.2\pm 2.7$ MeV, are much smaller
than expected from the leading SU(3)-breaking one-loop contribution.  It was
estimated that this source alone gives $\Delta_D^{\rm one~loop}\approx -47$
MeV and $\Delta_B^{\rm one~loop}\approx -16$ assuming that
$g_{D^{\ast}D\pi}^2=0.5$.  Using our values of $g_{D^{\ast}D\pi}$ and
$g_{B^{\ast}B\pi}$ instead would yield the less problematic numbers
$\Delta_D^{\rm one~loop}\approx -7$ MeV and $\Delta_B^{\rm one~loop}\approx
-3$.

\section*{Charge radii}

Gauge-invariance of the Lagrangian (\ref{chiral}) implies extra Feynman rules
for the matrix elements of the light-quark vector current between two mesons.
The photon can attach to the nonlocal vertices as well as to the quark lines.
The new Feynman rule for a vertex with outgoing photon momentum $q$ and
outgoing light-quark momentum $k$ is \begin{equation}
\frac{-i(2k+q)_{\mu}\gamma_5(\varepsilon\!\!\!/^{\ast})Z^2}{[-k^2+
\widehat{\Lambda}^2] [-(k+q)^2+ \widehat{\Lambda}^2] } .\end{equation} This
must be included in the computation of the light-quark contribution to the
electromagnetic charge radius of the meson (although it did not appear in the
above calculation of the $DD^{\ast}\gamma$ coupling).

Define the matrix element of the electromagnetic current by
\begin{equation}\langle H_Q(p^{\prime}) | J_{\mu}^{\rm e.m.} |H_Q(p)
\rangle = [-e_q F_q(q^2)+e_Q F_Q(q^2)](p+p^{\prime})_{\mu} ,\end{equation}
where
$e_c=e_u=2/3$ and
$e_b=e_d=-1/3$ and we have decomposed the form factor into light-quark and
heavy-quark contributions with $F_q(0)=1=F_Q(0)$. We define $r_i=[6 \partial
F_i/\partial q^2(0)]^{1/2}$ for $i=q,Q$.  As an example we compare the
contributions to the charge radius of the ground state pseudoscalar ($L=0$)
$D$ to one of the excited states treated in Part II, the scalar ($L=1$)
$D_0^{\ast}$.\begin{equation}
\begin{array}{ccc} H & 1/r_q({\rm MeV}) & 1/r_Q ({\rm MeV}) \\ \hline D & 300
& 830 \\ D_0^{\ast} & 690 & 830
\end{array} \end{equation}

The 300 MeV vs. 690 MeV shows that the excited state is more compact than the
ground state, as might be expected due to the higher energy present in the
light-quark system in the excited state. This effect is not present in models
of nonrelativistic quarks.

We also note that the large light-quark charge radius of the $D$ reflects a
rapid damping of the form factor in $q^2$ when the current attaches to the
light-quark line.  We will use this fact when discussing nonresonant
contributions to semileptonic decays in which a pion attaches to the
light-quark line.

\section*{Nonresonant processes}

The data indicate that the decays $\overline{B}^0\rightarrow
D^+\ell\overline{\nu}$ and $\overline{B}^0\rightarrow
D^\ast(2010)^+\ell\overline{\nu}$  contribute only 66\% of the total
semileptonic width $\overline{B}^0\rightarrow \ell\overline{\nu}+{\rm
anything}$. We will find in Part II that resonant decays to the eight lowest
lying excited $D$'s account for a further 6\%. This leaves 28\% missing.  It
is safe to assume that this will not be made up by decays into even higher
lying $D$ excitations.  And neither will it be decays to final states other
than charm, in view of the smallness of $|V_{ub}|$.  The implication is that
a significant portion of the total width must be into nonresonant charmed
final states.  The nonresonant width must be of order
$1.6\times10^{13}|V_{cb}|^2$ s$^{-1}$.

We consider nonresonant contributions to the processes $\overline{B}^0
\rightarrow D^{(\ast)}\pi\ell
\overline{\nu}$ in our model. The Lagrangian (\ref{chiral}) gives rise to
three graphs, one with a pion coming off the light-quark line (proportional
to the common mass $m$) and two with pions coming off the vertices.  The sum
of these three graphs is the same as an insertion of $f_{\pi}^{-1}
l_{\mu}\gamma^{\mu}\gamma_5$ on the light-quark line, where $l$ is the pion
momentum.  Because of the difficulties involved in computing the resulting
four-point graph, we were able only to obtain the hadronic matrix element in
the soft-pion limit, i.e. keeping only terms linear in $l$.  In this limit,
we define form factors $q_i(\omega)$ and $r_i(\omega)$ (with $\omega=v\cdot
v^{\prime}$ as usual) by \begin{eqnarray} \lefteqn{ f_{\pi}\langle D^0(M_D
v^{\prime})\pi^+(\l)|\overline{c}\gamma_{\mu}(1-\gamma_5)b
|\overline{B}^0(M_B v) \rangle = } \\  &&q_V
i\varepsilon_{\mu\nu\rho\sigma}v^{\nu}l^{\rho} v^{\prime \sigma}-q_{A_1}
l_{\mu} -[q_{A_2}l\cdot v +q_{A_3}l\cdot v^{\prime}] v_{\mu} -[q_{A_4}l\cdot
v +q_{A_5}l\cdot v^{\prime}] v^{\prime}_{\mu} .\nonumber \end{eqnarray} and
\begin{eqnarray} \lefteqn{ f_{\pi}\langle D^{\ast 0}(M_{D^{\ast}}
v^{\prime})\pi^+(\l)|\overline{c}\gamma_{\mu}(1-\gamma_5)b
|\overline{B}^0(M_B v) \rangle = } \nonumber \\  && \varepsilon \cdot v [
r_{V_1} l_{\mu} +(r_{V_2}v \cdot l +r_{V_3} v^{\prime}\cdot l) v_{\mu}
+(r_{V_4} v\cdot l +r_{V_5} v^{\prime}\cdot l) v^{\prime}_{\mu} ]  \nonumber
\\ &&+(r_{V_6} v\cdot l +r_{V_7} v^{\prime}\cdot l) \varepsilon_{\mu}
+(r_{V_8}v_{\mu}+r_{V_9}v^{\prime}_{\mu})l \cdot \varepsilon \nonumber \\ &&
-i \varepsilon_{\mu\nu\rho\sigma} \varepsilon^{\nu} [ r_{A_1}v^{\prime\rho}
v^{\sigma} + l^{\rho}(r_{A_2}v^{\sigma}+r_{A_3}v^{\prime\sigma})].
\end{eqnarray} The values of the above form factors at $\omega=1$ are
\begin{equation}\begin{array}{cccccc} q_V & q_{A_1} & q_{A_2} & q_{A_3} &
q_{A_4} & q_{A_5} \nonumber \\ \hline 14.8 & 24.0 & -3.5 & -8.2 & -4.6 & -2.7
\end{array} \end{equation} and
\begin{equation}\begin{array}{ccccc} r_{V_1} &
r_{V_2}+r_{V_3}+r_{V_4}+r_{V_5} &r_{V_6}+r_{V_7} &r_{V_8}+r_{V_9}
&r_{A_2}+r_{A_3} \nonumber \\ \hline -1.7 & 11 & -0.1 & -20.1 & 19.5 .
\end{array} \end{equation}

The rather large magnitude of some of these dimensionless hadronic quantities
is striking.  The amplitude vanishes for vanishing pion momenta, and it
rapidly increases for increasing pion momentum up to some momenta of order
300 or 400 MeV (roughly the inverse of the charge radius), after which the
amplitude will decrease. We choose a small value of the pion momentum (equal
to the pion mass) well below where the amplitude peaks.   We set
$v=v^{\prime}$, let the massless electron and antineutrino emerge with equal
energies, and then perform the four-body phase space integral with this
constant matrix element squared. The result is
$\Gamma(\overline{B}^0\rightarrow D^0\pi^+\ell \overline{\nu})\sim 0.6
\times10^{13}|V_{cb}|^2$ s$^{-1}$ and
$\Gamma(\overline{B}^0\rightarrow D^{\ast 0}\pi^+\ell
\overline{\nu})\sim 0.3 \times10^{13}|V_{cb}|^2$ s$^{-1}$.\footnote{These
results are also quite sensitive to the quark masses used.  For example, if
our ``new'' values of the masses are used then
$\Gamma(\overline{B}^0\rightarrow D^0\pi^+\ell \overline{\nu})$ increases by
40\%.}  The decays with charged
$D$ and neutral pion are half as large. The total nonresonant width is then
$\sim 1.8\times10^{13}|V_{cb}|^2$ s$^{-1}$, in rough agreement with the value
deduced above. While this crude estimate is far from being a definite
prediction, it does make reasonable the possibility that nonresonant decays
make an important contribution to the semileptonic width of the $B$ meson.

\section*{Part II} \section*{Description of model for excited mesons}

Our notation is $N\;J^P_{j_{\ell}}$, where $N$ labels the radial excitation,
$J^P$ is the meson spin-parity, and $j_{\ell}$ is the total angular momentum
of the light degrees of freedom.  It is the sum of orbital angular momentum
$L$ and light-quark spin $s_{\ell}=1/2$.  This paper will treat the eight
$N=1$ states shown in (\ref{states}), where the masses in the last column
corresponding to question marks are estimates. The naming convention is that
of the Particle Data Group \cite{pdg}. \begin{equation} \begin{array}{cccccc}
N & L & j_{\ell} &       {\rm doublet}  & {\rm name} & {\rm mass}  \\ \hline
1 & 1 & 1/2     & \left(\begin{array}{c} 1\;0^-_{1/2} \\
1\;1^-_{1/2}\end{array}\right) & \begin{array}{c} D \\ D^{\ast}(2010)
\end{array} & \begin{array}{c} 1869  \\  2010 \end{array}\\  1 & 1 &   1/2 &
\left(\begin{array}{c} 1\;0^+_{1/2} \\ 1\,1^+_{1/2} \end{array}\right) &
\begin{array}{c} D_0^{\ast}(?) \\ D_1(?) \end{array} & \begin{array}{c}  2465
\\  2270 \end{array}\\ 1 & 1 &   3/2     & \left(\begin{array}{c}
1\;1^+_{3/2} \\ 1\;2^+_{3/2}\end{array}\right) & \begin{array}{c} D_1(2420)
\\ D_2^{\ast}(2460) \end{array} & \begin{array}{c} 2421  \\  2465
\end{array}\\ 1 & 2 &   3/2     & \left(\begin{array}{c} 1\;1^-_{3/2} \\
1\;2^-_{3/2} \end{array}\right) & \begin{array}{c} D^{\ast}(?) \\ D_2(?)
\end{array} & \begin{array}{c} 2800  \\  2800 \end{array} \end{array}
\label{states} \end{equation} In addition, we will consider the first radial
excitation $2\;0^-_{1/2}$ with an estimated mass of 2440 MeV. As is commonly
done \cite{cleoexc}, we will ignore possible mixing between $1\;1^+_{1/2}$ and
$1\;1^+_{3/2}$.

The interaction vertices between hadron, heavy quark, and light degrees of
freedom are chosen in the following way.  Each vertex is the product of some
gamma-matrix structure and a damping factor.  The gamma-matrix structure is
chosen to be identically equal to the form determined by heavy-quark symmetry
\cite{falk}.  This is required for consistency with QCD in the heavy-quark
limit.  The damping factor is chosen to be a sum of terms of the form
$[Z^2/(-k^2+\widehat{\Lambda}^2)]^n$, where the least power $n$ is chosen to
be the smallest which guarantees the convergence of the relevant integrals,
and where $Z$ and $\widehat{\Lambda}$ are different for each state. We use
standard quark propagators throughout.

We take the interaction Lagrangian to be that given in (\ref{ww}) where ${\cal
L}_H(x,y)$ and the powers $n$ are given by \begin{equation}
\begin{array}{ccc} H & {\cal L}_H(x,y) & n \\ \hline 1\;0^-_{1/2} & -iH(x)
\gamma_5 q(y) & 1\\ 1\;1^-_{1/2} & iH\!\!\!\!/(x) q(y) & 1\\ 1\;0^+_{1/2} &
iH(x) q(y) & 1\\ 1\;1^+_{1/2} & iH\!\!\!\!/(x) \gamma_5 q(y) & 1\\
1\;1^+_{3/2} &
\frac{1}{\widehat{\Lambda}}\sqrt{\frac{3}{2}} H_{\mu}(x) \gamma_5 \left\{
g^{\mu\nu}-\frac{1}{3}\gamma^{\mu} ( \gamma^{\nu} -v^{\nu})   \right\}
\partial_{\nu}q(y) & 2\\ 1\;2^+_{3/2} & \frac{1}{\widehat{\Lambda}}
H^{\mu\nu}(x) \gamma_{\mu}\partial_{\nu} q(y)  & 2\\ 1\;1^-_{3/2} &
\frac{1}{\widehat{\Lambda}}\sqrt{\frac{3}{2}}H_{\mu}(x) \left\{
g^{\mu\nu}-\frac{1}{3}\gamma^{\mu} ( \gamma^{\nu} +v^{\nu})   \right\}
\partial_{\nu}q(y) & 2\\ 1\;2^-_{3/2} & \frac{1}{\widehat{\Lambda}}
H^{\mu\nu}(x) \gamma_{\mu}\gamma_5 \partial_{\nu} q(y) & 2 \\ 2\;0^-_{1/2} &
-iH(x) \gamma_5 q(y) & 1+2. \end{array} \label{vertices}.
\end{equation} Wherever a velocity $v$ occurs in (\ref{vertices}), it is the
velocity of the external meson (and not that of the heavy quark).

The only parameters of the model are the quark masses; the various $Z$'s and
$\widehat{\Lambda}$'s are fixed via the meson ``mass functions"
$\Sigma(p^2)$.   These are defined in terms of the meson self-energy graphs,
which are given by $i\Sigma(p^2)$, $-ig_{\mu\nu}\Sigma(p^2)+\cdots$ and
$ig_{\mu\nu}g_{\rho\sigma}\Sigma(p^2)+\cdots$ when $J=0,1$ and 2,
respectively. First $\widehat{\Lambda}$ is fixed by setting $\Sigma(M^2)=0$.
Then, $Z$ is fixed by setting $\Sigma^{\prime}(M^2)=1$, where prime denotes
differentiation with respect to $p^2$.  This latter condition is equivalent
to conservation of the heavy-quark vector current by the Ward identity.

The situation is a little more complicated in the case of the radial
excitation $2\;0^-_{1/2}$ of the ground-state pseudoscalar $1\;0^-_{1/2}$. We
take the damping factor to be a linear combination of $n=1$ and $n=2$, which
we write as \begin{equation} \frac{Z^2}{-k^2+\widehat{\Lambda}^2}\left[1 + f
\frac{\widehat{\Lambda}^2}{-k^2+\widehat{\Lambda}^2} \right] .\end{equation}
First, $f$ and $Z$ are determined as functions of $\widehat{\Lambda}$ at
fixed quark and meson masses by the zero and the slope of the mass function.
Then, $\widehat{\Lambda}$ is determined by the vanishing of the matrix
element $\langle 1\;0^-_{1/2}|\overline{Q}\gamma_{\mu}Q | 2\;0^-_{1/2}
\rangle$ at $q^2=0$,  as required by current conservation.

The values of the various constants are given in (\ref{LsandZs}) in MeV.
Quark masses $m_c=1440$ and $m_q=250$ MeV have been used.

\begin{equation} \begin{array}{ccc} H & \widehat{\Lambda}_H & Z_H \\
\hline 1\;0^-_{1/2} & 537 & 745 \\ 1\;1^-_{1/2} & 762 & 1193 \\ 1\;0^+_{1/2}
& 1217 & 1838 \\ 1\;1^+_{1/2} & 1065 & 1881 \\ 1\;1^+_{3/2} & 1400 & 2143 \\
1\;2^+_{3/2} & 1359 & 2122 \\ 1\;1^-_{3/2} & 1976 & 3231
\\ 1\;2^-_{3/2} & 1745 & 2804 \\ 2\;0^-_{1/2} & 1430 & 2594 \end{array}
\label{LsandZs}
\end{equation} We find $f=-1.1$ for the $2\;0^-_{1/2}$ state.  In addition,
we need the values appropriate to the $B$ and $B^{\ast}$: $m_b=4800$,
$\widehat{\Lambda}_B=621$, $\widehat{\Lambda}_{B^\ast}=702$, $Z_B=1103$, and
$Z_{B^\ast}=1290$ MeV.

\section*{The resonant processes $\overline{B}\rightarrow
H\ell\overline{\nu}$}

We now consider the hadronic matrix elements \begin{equation} {\cal
M}_{\mu}^H = \frac{1}{\sqrt{M M^{\prime}}}\langle H(M^{\prime}v^{\prime})
|\overline{c}\gamma_{\mu}(1-\gamma_5)b|\overline{B}(M v) \rangle ,
\label{hadmtxelt} \end{equation} where $H$ denotes any of the
$D$ states. The matrix elements may be written as

\begin{equation} \begin{array}{ll}  H & {\cal M}_{\mu}^H \\ \hline
1\;0^-_{1/2} & S(a_+,a_-)_{\mu} \\ 1\;1^-_{1/2} &
T(b_V,b_{A_1},b_{A_2},b_{A_3})_{\mu\nu} \varepsilon^{\ast\nu}  \\
1\;0^+_{1/2} & -S(c_+,c_-)_{\mu} \\ 1\;1^+_{1/2} &
-T(d_A,d_{V_1},d_{V_2},d_{V_3})_{\mu\nu}\varepsilon^{\ast \nu} \\
1\;1^+_{3/2} &  -T(f_A,f_{V_1},f_{V_2},f_{V_3})_{\mu\nu} \varepsilon^{\ast
\nu} \\  1\;2^+_{3/2} & T(g_V,g_{A_1},g_{A_2},g_{A_3})_{\mu\nu}
\varepsilon^{\ast \nu\lambda}v_{\lambda} \\ 1\;1^-_{3/2} &
T(h_V,h_{A_1},h_{A_2},h_{A_3})_{\mu\nu}\varepsilon^{\ast \nu} \\ 1\;2^-_{3/2}
& -T(k_A,k_{V_1},k_{V_2},k_{V_3})_{\mu\nu}\varepsilon^{\ast
\nu\lambda}v_{\lambda} \\ 2\;0^-_{1/2} & S(l_+,l_-)_{\mu}, \end{array}
\label{mtxelts} \end{equation} where \begin{eqnarray} S(a_+,a_-)_{\mu} & = &
a_+ (v+v^{\prime})_{\mu} + a_- (v-v^{\prime})_{\mu} \nonumber \\
T(b_V,b_{A_1},b_{A_2},b_{A_3})_{\mu\nu} &=& ib_V
\varepsilon_{\mu\nu\rho\sigma}  v^{\prime\rho} v^{\sigma} - b_{A_1}g_{\mu\nu}
-(b_{A_2}v_{\mu}+b_{A_3}v^{\prime}_{\mu})v_{\nu}. \label{mtxeltforms}
\end{eqnarray}  We take the form factors to be functions of $\omega=v\cdot
v^{\prime}$; e.g. $a_+\equiv a_+(\omega)$.

The form factors $a_{\pm}$ and $b_i$ of (\ref{mtxelts}) and
(\ref{mtxeltforms}) become proportional to the ``original" Isgur-Wise
function $\xi\equiv \xi_{1/2}$ in the heavy-quark limit
$m_{b,c}\rightarrow\infty$. Similarly, $c_{\pm}$ and $d_i$ become
proportional to $\tau_{1/2}$, $f_i$ and $g_i$ to $\tau_{3/2}$, and $h_i$ and
$k_i$ to $\xi_{3/2}$. The form factors $l_{\pm}$ for decay into the radially-
excited state become proportional to $\xi^{(2)}_{1/2}$.   The coefficients of
proportionality are shown in (\ref{hqformfacs}).

\begin{equation} \begin{array}{l|ll} {\rm units} & {\rm
form\;factors\;in\;given \;units} & \\ \hline \xi_{1/2} & a_+=1 & a_-=0 \\ &
b_V=1 & b_{A_1}=\omega+1 \\ & b_{A_2}=0 & b_{A_3}=-1 \\ \hline \tau_{1/2} &
c_+=0 & c_-=-2    \\ & d_{V_1}=2(\omega-1) & d_{V_2}=0 \\ & d_{V_3}=-2 &
d_A=2  \\ \hline \tau_{3/2}   &  f_{V_1}=-\frac{\omega^2-1}{\sqrt{2}}  &
f_{V_2}=-\frac{3}{\sqrt{2}}  \\ & f_{V_2}-f_{V_3}=-\frac{\omega+1}{\sqrt{2}}
&  f_A= -\frac{\omega+1}{\sqrt{2}} \\ &  g_V=\sqrt{3}  &
g_{A_1}=\sqrt{3}(\omega+1)  \\ &  g_{A_2}=0  & g_{A_3}=-\sqrt{3}  \\  \hline
\xi_{3/2}  &  h_V=\frac{\omega-1}{\sqrt{2}}  &
h_{A_1}=\frac{\omega^2-1}{\sqrt{2}}  \\ &  h_{A_2}=\frac{3}{\sqrt{2}}  &
h_{A_2}+h_{A_3}=-\frac{\omega-1}{\sqrt{2}}  \\ &  k_{V_1}=\sqrt{3}(\omega-1)
&  k_{V_2}=0   \\ &  k_{V_3}=-\sqrt{3}  & k_A=\sqrt{3}  \\ \hline
\xi^{(2)}_{1/2}  &  l_+=\omega-1  &  l_-=0 \end{array} \label{hqformfacs}
\end{equation}

Our results for the various full-model form factors for $B$ decays into the
excited $D$ states are represented in Figs. 2-4.  In order to facilitate
comparison with the heavy-quark limit, we have chosen not to plot the
full-model form factors directly, but rather to divide out the coefficients
in (\ref{hqformfacs}).  This puts all form factors on a comparable scale, and
shows the ``effective" Isgur-Wise functions which would be extracted from the
data under the assumption that the deviations from the heavy-quark limit were
small.  In those cases where the coefficient vanishes at $\omega=1$, we have
subtracted off the zero-recoil value before dividing by the coefficient.  For
example, the curve labelled ``$d_{V_1}$" in Fig. 2 is equal to
$[d_{V_1}(\omega)-d_{V_1}(1)]/2(\omega-1)$.  In those cases where the
coefficient vanishes identically, we have omitted the corresponding
full-model form factor for brevity.

Fig. 5 shows a comparison between the full-model form factors for $B$ decays
into the ground state pseudoscalar $D$ and its first radial excitation.  We
note the strong suppression of the form factors for the latter.

The various decay widths are given by \begin{equation} \Gamma_H = |V_{cb}|^2
\frac{G_F^2 M^5 r^3}{48 \pi^3} \int_{1}^{(1+r^2)/2r} d\omega
\sqrt{\omega^2-1} {\cal F}_H(\omega), \label{wspec} \end{equation} where
$r=M^{\prime}/M$.  The functions ${\cal F}_H$ are given by

\begin{equation} \begin{array}{ll} H & {\cal F}_H(\omega) \\ \hline
1\;0^-_{1/2} & F_0(a_+,a_-) \\  1\;1^-_{1/2} & 2F_T(b_V,b_{A_1}) +
F_L(b_{A_1},b_{A_2},b_{A_3}) \\  1\;0^+_{1/2} & F_0(c_+,c_-) \\  1\;1^+_{1/2}
& 2F_T(d_A,d_{V_1}) + F_L(d_{V_1},d_{V_2},d_{V_3}) \\ 1\;1^+_{3/2} &
2F_T(f_A,f_{V_1}) + F_L(f_{V_1},f_{V_2},f_{V_3}) \\ 1\;2^+_{3/2} &
(\omega^2-1)\{F_T(g_V,g_{A_1}) + \frac{2}{3}F_L(g_{A_1},g_{A_2},g_{A_3}) \}
\\ 1\;1^-_{3/2} & 2F_T(h_V,h_{A_1}) + F_L(h_{A_1},h_{A_2},h_{A_3}) \\
1\;2^-_{3/2} & (\omega^2-1)\{F_T(k_A,k_{V_1}) +
\frac{2}{3}F_L(k_{V_1},k_{V_2},k_{V_3}) \} \\ 2\;0^-_{1/2} & F_0(l_+,l_-) ,
\end{array} \label{wargs} \end{equation} where \begin{eqnarray} F_0(a_+,a_-)
&= &(\omega^2-1)[(1+r)a_+-(1-r)a_-]^2  \\ \nonumber F_T(b_V,b_{A_1}) &=&
(1-2r\omega+r^2)[(\omega^2-1)b_V^2+b_{A_1}^2] \\ \nonumber
F_L(b_{A_1},b_{A_2},b_{A_3}) &=&[(\omega-r)b_{A_1} +
(\omega^2-1)(rb_{A_2}+b_{A_3})]^2. \label{wsubargs} \end{eqnarray} We compare
our model results for the full rates $\Gamma_H$ with those of other models in
(\ref{comparison}), in units of $10^{13}|V_{cb}|^2$ s$^{-1}$.   (Note that
$1^+_{1/2}$ and $1^+_{3/2}$ are linear combinations of $^1P_1$ and $^3P_1$.
Results computed in the latter basis are denoted by asterisks in
(\ref{comparison}), and ``HQ" denotes heavy-quark limit results.)

\begin{equation}  \begin{array}{cc|llcllll}

N & J^P_{j_{\ell}} & {\rm ISGW} & {\rm CNP} &  {\rm ours}  & {\rm ours} &
{\rm CNP} & {\rm Suzuki} \\ &  &{\rm (full)} &{\rm (full)} & {\rm (full)} &
{\rm (HQ)} &  {\rm (HQ)} &  {\rm (HQ)}\\ &        &\cite{daryl}&\cite{CNP1}&
& &\cite{CNP2}&\cite{suzuki}
\\ \hline 1 &  0^-_{1/2} & 1.25  & 0.76   & 0.900 & 0.771 & 0.723 & 0.741 \\
1 &  1^-_{1/2} & 3.24  & 2.3     & 2.91  & 2.32  & 2.22  & 2.20 \\  1 &
0^+_{1/2} & 0.049 & 0.076   & 0.028 & 0.013 & 0.026 & 0.011  \\ 1  &
1^+_{1/2} & 0.180^*  &       & 0.040 & 0.026 & 0.036 & 0.015  \\  1 &
1^+_{3/2} & 0.045^* & 0.076^*   & 0.117 & 0.070 & 0.052 & 0.034  \\ 1 &
2^+_{3/2} & 0.061 &         & 0.076 & 0.104 & 0.103 & 0.049  \\   1 &
1^-_{3/2} &       &         & 0.0021& 0.0002&       & 0.0007  \\   1  &
2^-_{3/2} &       &       & 0.0003& 0.0002&       & 0.0007  \\   2 &
0^-_{1/2} & 0.013 &        & 0.055 &       &       & 0.045   \\   2 &
1^-_{1/2}  & 0.010 &        &       &       &       & 0.115

\end{array} \label{comparison} \end{equation} There are clear similarities
among the various models.  At the same time, measurements of these exclusive
resonant processes could potentially discriminate between the models.

In Fig. 6 we show the electron energy spectra for the various resonant final
$D$ states and their total contribution to the inclusive $B$ decay spectrum.
The location of the maximum agrees with ISGW \cite{daryl}. Fig. 7 shows the
spectra of the excited $D$ final states in more detail.

\section*{Part III}
\section*{Some details of the heavy-quark limit.}

We first recover one of the basic properties of heavy quark effective
theory.  The following is satisfied identically: \begin{equation} \langle P |
\overline{Q}(x)(i\partial\!\!\!/-m_Q)Q(x) | P \rangle =0 \label{eom}
\end{equation} when the pseudoscalar meson $P$ is on shell.  This is easily
seen because $i\partial\!\!\!/-m_Q$ just removes one of the heavy-quark
propagators from the graph for (\ref{eom}), leaving the $P$ mass function
evaluated at $p^2=M^2$.  This vanishes, by the definition of the mass.  In
the heavy-quark effective theory at lowest order we have $Q(x)\simeq
\exp(-im_Qv\cdot x) h_v(x)$. Using
$v\!\!\!/ h_v= h_v$, we find \begin{equation} \langle P |
\overline{Q}(x)(i\partial\!\!\!/-m_Q)Q(x) | P \rangle \simeq \langle P |
\overline{h}_v(x)i\partial\cdot v h_v(x) | P \rangle = 0 \label{eomexp}
.\end{equation}  Thus we recover the equation of motion $iv\cdot D h_v =0$ of
heavy quark effective theory; here $D=\partial$ since there are no gluons in
the model.

The heavy-quark limit results in (\ref{comparison}) are included for
comparison purposes only.  They have limited physical meaning since they are
computed using the above formulae with Isgur-Wise functions for the form
factors, while the meson masses are kept at their physical values.

It has become standard to display the differential $\omega$ spectrum by
taking the square root of (\ref{wspec}) and dividing out model-independent
factors in such a way as to make the result equal to the corresponding
Isgur-Wise function if the form factors are replaced by their values in the
heavy-quark limit. This procedure is applied to the $1/2^+$ and $3/2^+$ final
states (the curves labelled ``full") in Figs. 8 and 9, respectively.  We note
the considerable deviations from the heavy-quark limit.

An important quantity is the lowest-order mass difference
$\overline{\Lambda}$ between meson and heavy quark: \begin{equation}
M=m_Q+\overline{\Lambda} + {\cal O}(\overline{\Lambda}^2/m_Q) .
\label{Lbar}
\end{equation}   We extract the following approximate values for the
$N=1$ states:
\begin{equation} \begin{array}{ccccc} j_{\ell}^P & 1/2^- & 1/2^+ & 3/2^+ &
3/2^- \\ \hline  \overline{\Lambda} & 500 & 860 & 990 & 1330 \end{array}
\label{LLbarvals} \end{equation}  These numbers make clear the breakdown of
an expansion in
$\overline{\Lambda}/m_c$ for the higher-lying states.

We also display the following zero-recoil values of the Isgur-Wise functions,
along with our previous result
\cite{linear} for the slope of the usual Isgur-Wise function $\xi\equiv
\xi_{1/2}$ at zero recoil.
\begin{equation} \begin{array}{cccc} -\xi_{1/2}^{\prime}(1) & \tau_{1/2}(1) &
\tau_{3/2}(1) & \xi_{3/2}(1) \\ \hline 1.28 &  0.21 & 0.29 & 0.013
\end{array} \label{IWvals} \end{equation} In comparison, the ISGW model
result is $\tau_{1/2}(1) = \tau_{3/2}(1) = 0.315$ \cite{IWcharm}.  The QCD
sum rule approach of \cite{CNP2} found $\tau_{1/2}(1) \simeq \tau_{3/2}(1) =
0.2\;{\rm to}\;0.25$,\footnote{The range of values reflects the conflict
between Fig. 1 and Table 1 of \cite{CNP2}.} while the Bethe-Salpeter approach
of \cite{DHJ} found $\tau_{1/2}(1)=0.21$ and $\tau_{3/2}(1)=0.44$.

Another issue is the extent to which the Bjorken sum rule is saturated by the
states we have calculated.  This sum rule reads \cite{IWcharm}
\begin{equation} -\xi_{1/2}^{\prime}(1) = 1/4 + |\tau_{1/2}(1)|^2 + 2
|\tau_{3/2}(1)|^2  + \ldots  \label{bjsr} \end{equation} where the ellipsis
denotes contributions from higher radial excitations, from states with
quantum numbers other than $1/2^-$, $1/2^+$ and $3/2^+$,  and from inelastic
continua. We find numerically $1.28=0.25+0.04+0.17+\ldots=0.46+\ldots$ for
Eq. (\ref{bjsr}).  The sum rule is far from being saturated by the resonances
we have considered so far; this result reflects the situation already
observed in the full model.  As already stated, our model indicates that this
is due mainly to nonresonant contributions to final states containing a pion.
The discrepancy is even larger here because the sum rule was derived in the
heavy-quark limit, and the $D$ mesons are rather far from this limit in our
model.

One last issue of current interest is the dependence of the slope of the
Isgur-Wise function on the light-quark mass.  We find that the slope
increases when the light-quark mass is replaced by the strange-quark mass,
from -1.28 to -1.55.  This behavior agrees with other analyses in quark
models \cite{close}, sum rules \cite{huang} and lattice \cite{lat}, but it
disagrees with heavy-quark chiral perturbation theory \cite{jenkins}.  Our
dependence of the slope on the light-quark mass is given in
\cite{lambdabaryons}, and it is far from linear.

\section*{The decay $D\rightarrow K^\ast \ell\nu$}

We now explore how the model may be extended to describe strange mesons. We
will consider the $K^\ast$ and will avoid the $K$ and its complicating
pseudo-Goldstone nature.   The process $D^+\rightarrow
\overline{K}^\ast(892)^0 e^+\nu_e$ is of interest since models tend to
disagree with the data by predicting a branching fraction which is larger
than the measured value.   For us, one option is to model the $K^\ast$
exactly like we model the $D^\ast$ and $B^\ast$ mesons.  This would be making
the rather suspect assumption that the $s$ quark acts like a heavy quark. The
other option, perhaps equally suspect, is to treat the $s$ quark as light, on
the same footing as the $u$ and $d$ quarks.  This would entail modifying the
vertex damping factor in (\ref{damping}) in order to treat the $\overline d$
and $s$ symmetrically. The damping factor then becomes \begin{equation}
\frac{Z^2}{-k_{\overline d}^2+\widehat{\Lambda}^2} +
\frac{Z^2}{-k_s^2+\widehat{\Lambda}^2}. \end{equation} We will give the
results for these two options.

With a strange quark mass of $250+170=420$ MeV, we find
$\widehat{\Lambda}_{K^\ast} = 598$ MeV and $Z_{K^\ast} = 836$ MeV. The
relevant observables can be compared to the measured values as
follows.\cite{pdg} \begin{equation}\label{DKstar} \begin{array}{cccc} &
\Gamma_L/\Gamma_T  & \Gamma_+/\Gamma_- & {\rm B}(D^+ \rightarrow
\overline{K}^{\ast 0}e^+\nu_e)\% \\ \hline {\rm experiment} & 1.23 \pm 0.13 &
0.16 \pm 0.04 & 4.8 \pm 0.4 \\ {\rm model~(heavy~}s) & 0.99 & 0.24 & 8.1 \\
{\rm model~(light~}s) &1.04 & 0.25 & 2.8 \\ \end{array} \end{equation} We see
that only the branching fraction is very sensitive to how the $s$ quark is
treated, and interestingly enough, the experimental branching fraction lies
between the two extreme ways of treating the $s$.

We may also extract the form factors $V(q^2)$, $A_1(q^2)$ and $A_2(q^2)$ for
$q^2$=0, which may then be compared to nonrelativistic quark model and to sum
rule results.  For additional models see \cite{pdg}. \begin{equation}
\begin{array}{c|ccc} & V(0) & A_1(0) & A_2(0) \\ \hline {\rm model~(heavy~}s)
& 1.14 & 0.76 & 0.82 \\ {\rm model~(light~}s) & 0.68 & 0.48 & 0.52 \\ {\rm
ISGW} & 1.1 & 0.8 & 0.8 \\ {\rm BBD} & 1.1 & 0.5 & 0.6 \\ {\rm experiment} &
1.1 \pm 0.2 & 0.56 \pm 0.04 & 0.40 \pm 0.08 \\ \end{array} \end{equation}
Note that the experimental values here \cite{pdg} have been extracted by
assuming specific pole expressions for the form factors.  We may compare those
form factors to our model form factors by fitting our form factors in the
physical region to the pole form $F({q}^{2}) =F(0)/(1-{q}^{2}/{m}_{F}^{2}
)$.  The values for ${m}_{F}$ in GeV are the following.
\begin{equation}\begin{array}{c|ccc}&V&{A}_{1}&{A}_{2}\\ \hline {\rm
assumed~in~\cite{pdg}}&2.1&2.5&2.5\\ {\rm model~(heavy~}s)&1.65&2.3&1.8\\
{\rm model~(light~}s)&2.0&3.8&2.4\end{array}\end{equation}

\section*{The form of the vertex damping factor}

The model relies on a vertex damping factor to suppress the flow of large
Euclidean momenta through light-quark propagators, and a specific pole form
has been chosen.\begin{equation} F(k^2) =
{\frac{{Z}^{2}}{-{k}^{2}+{\widehat{\Lambda}}^{2}}}\label{ee}
\end{equation}  This introduces singularities into our quark loop amplitudes
for momenta outside the physical region
\cite{anom}.  Viewing our amplitudes as functions of the quark masses while
keeping the meson masses fixed, there are also cusp-like singularities for
special (presumably unphysical) values of the quark masses
\cite{linear}.   We wish to explore here how our conclusions depend on our
choice of the damping factor, and on the singularity structure of the model in
particular.  To do this we will substitute an exponential form for the damping
factor,
\begin{equation} F(k^2) = R\exp\left(\frac{k^2}{\Delta^2}\right).
\end{equation} This may be less realistic for large
$-k^2$, but it is free from unphysical singularities at finite $-k^2$.  The
parameters $R$ and
$\Delta$ will be determined for each meson from the self-energy graphs, as
was done for the quantities $Z$ and $\widehat{\Lambda}$ of (\ref{ee}).  Our
standard quark masses imply $R_B$=3.81, $\Delta_B$=620 MeV, $R_D$=2.37,
$\Delta_D$=518 MeV, $R_{D^\ast}$=2.96 and
$\Delta_{D^\ast}$=785 MeV. The shape of the exponential damping factor is
compared to the original damping factor for the $B$ meson in Fig. 10.

We give the values of the two form factors for $B \rightarrow
D^{(\ast)}\ell\overline{\nu}$  which at $\omega$=1 are protected from first
order corrections in the heavy quark expansion.
\begin{equation}
\begin{array}{cc|cc} {\rm vertex} & m_c & b_{A_1}/2 & a_+ \\ \hline {\rm
original}  & 1.44 & 1.16 & 1.11 \\ & 1.45 & 1.12 & 1.17 \\ \hline {\rm
exponential} & 1.55 & .95 & 1.83 \\ & 1.50 & .97 & 1.34 \\ & 1.44 & 1.01 &
1.19 \\ & 1.40 & 1.06 & 1.17 \\ & 1.35 & 1.15 & 1.17\\ & 1.30 & 1.56 & 1.19
\\ \end{array}\label{exptable} \end{equation}  As we have explained elsewhere
\cite{linear}, the unphysical procedure of changing the quark masses while
holding the meson masses fixed can lead to clearly unphysical values for
amplitudes.  The interesting result in the exponential case is the very large
corrections to
$a_+$ for any reasonable value of $m_c$.  Thus our conclusion regarding the
existence of large corrections to the heavy-quark limit in at least one of $B
\rightarrow D^{\ast}\ell\overline{\nu}$ or $B \rightarrow D\ell\overline{\nu}$
\cite{hyperfine} remains unchanged.

\section*{Conclusions}

We have thoroughly explored a relativistic quark model for hadrons containing
one heavy quark.  It is extremely economical in its description of
nonperturbative QCD, its only parameters being the quark masses, yet it is
consistent with the symmetries of QCD.  It displays overall agreement, across
a broad range of different processes, with the data when it exists and with
more complicated models such as QCD sum rules in the absence of data.  The
main difference with other models is the suggestion that large
nonperturbative departures from the heavy quark limit occur in certain
quantities, most notably in the processes
$B\rightarrow D^{(\ast)}\ell \nu$ and in the heavy-quark contribution to the
$DD^\ast\gamma$ coupling.  At the same time, it is consistent with data for
the shape of the spectrum for
$B\rightarrow D^{\ast}\ell\nu$ and for the branching ratios for
$D^{\ast}\rightarrow D\gamma$.  To test the pattern of deviations from the
heavy-quark limit in our model and others, we especially encourage efforts to
extract and compare
$V_{cb}$ from each of $B\rightarrow D^{\ast}\ell\nu$, $B\rightarrow D\ell\nu$
and
$\Lambda_b \rightarrow \Lambda_c \ell \nu$.

\section*{Acknowledgements}

R.L. thanks M. Clayton for helpful comments and B.H. thanks M. Luke for
discussions.  This research was supported in part by the Natural Sciences and
Engineering Research Council of Canada.

\newpage \section*{Figure captions}

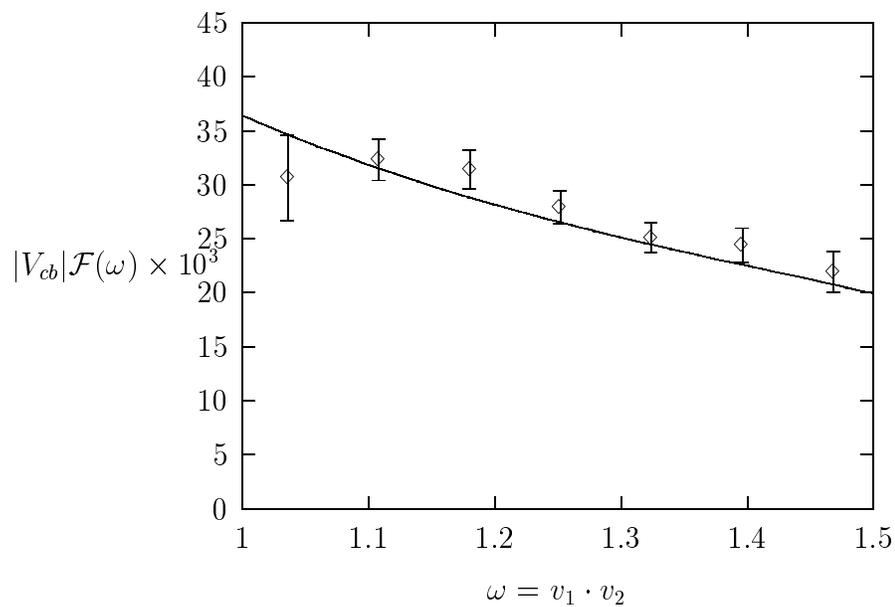
\noindent {\bf FIG. 1:} Comparison of model prediction for $B\rightarrow
D^{\ast}\ell\overline{\nu}$ with recent CLEO data \cite{cleo}. The notation
is the same as in \cite{cleo}. \vspace{4ex}

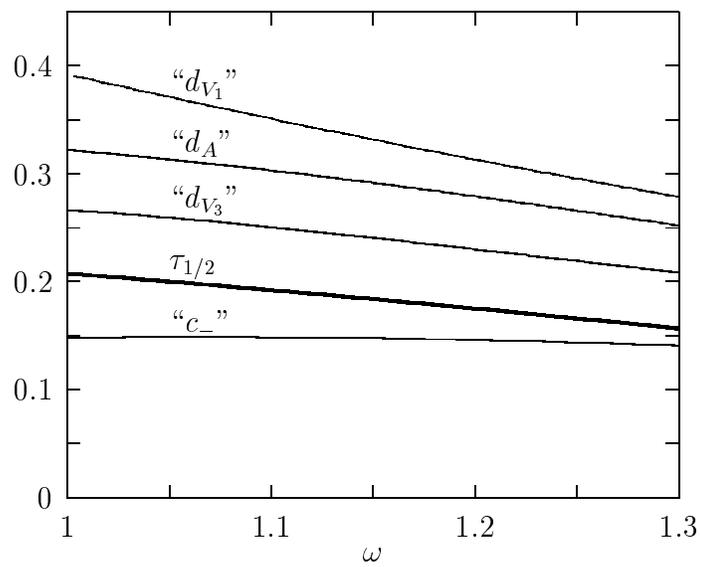
\noindent {\bf FIG. 2:}  Full-model form factors $c_i$ and $d_i$, defined in
(\ref{mtxelts}) and (\ref{mtxeltforms}).  Certain coefficients have been
divided out as explained in the text. Also shown is the Isgur-Wise function
$\tau_{1/2}$, defined in (\ref{hqformfacs}). \vspace{4ex}

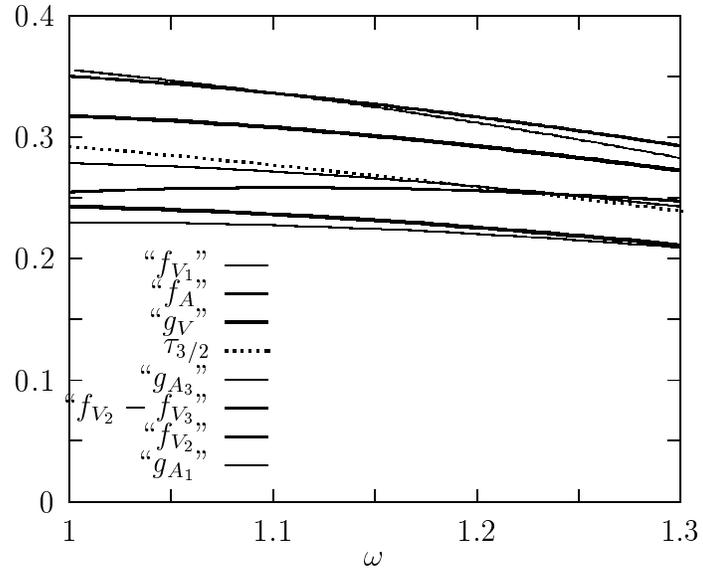
\noindent {\bf FIG. 3:} Full-model form factors $f_i$ and $g_i$, defined in
(\ref{mtxelts}) and (\ref{mtxeltforms}).  Certain coefficients have been
divided out as explained in the text. Also shown is the Isgur-Wise function
$\tau_{3/2}$, defined in (\ref{hqformfacs}). \vspace{4ex}

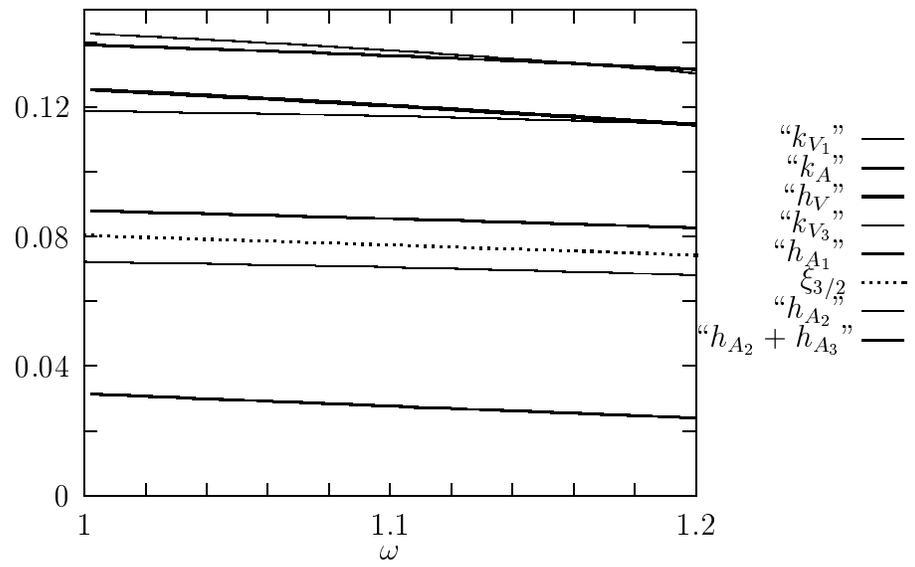
\noindent {\bf FIG. 4:} Full-model form factors $h_i$ and $k_i$, defined in
(\ref{mtxelts}) and (\ref{mtxeltforms}).  Certain coefficients have been
divided out as explained in the text. Also shown is the Isgur-Wise function
$\xi_{3/2}$, defined in (\ref{hqformfacs}). \vspace{4ex}

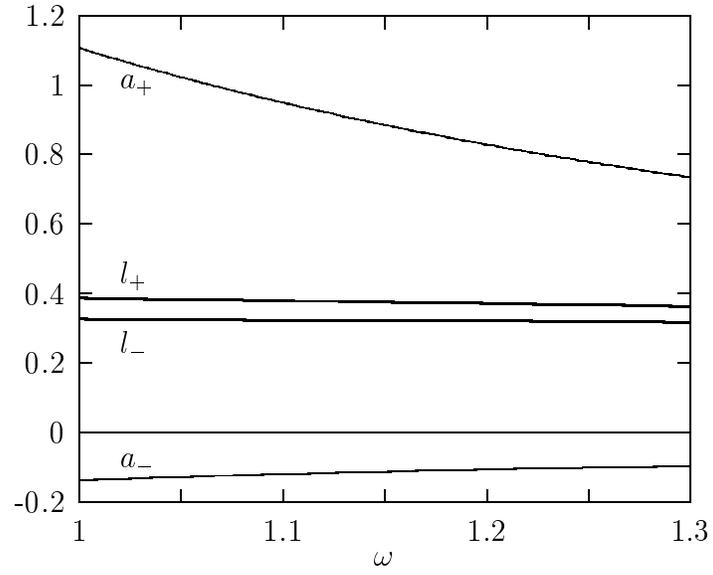
\noindent {\bf FIG. 5:} Full-model form factors for $B\rightarrow D\ell\nu$.
$a_{\pm}$ are for the ground-state $D$, and $l_{\pm}$ are for its first
radial excitation, as defined in (\ref{mtxelts}) and (\ref{mtxeltforms}).
\vspace{4ex}

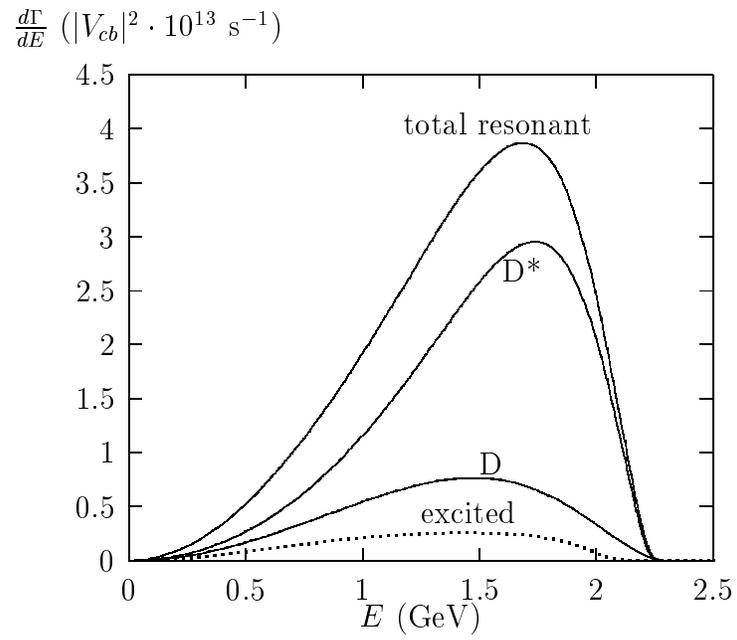
\noindent {\bf FIG. 6:} Electron energy spectra for resonant semileptonic $B$
decays in the full model. \vspace{4ex}

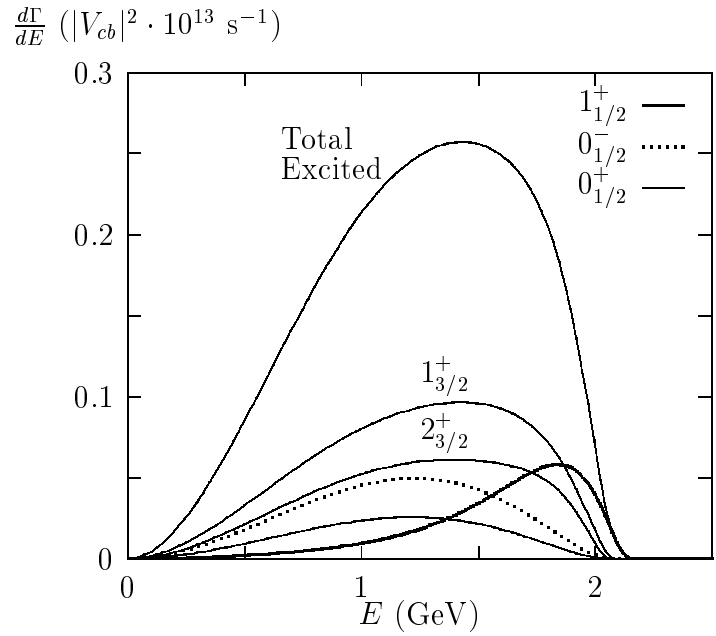
\noindent {\bf FIG. 7:} Electron energy spectra for resonant semileptonic $B$
decays to excited $D$ states only. \vspace{4ex}

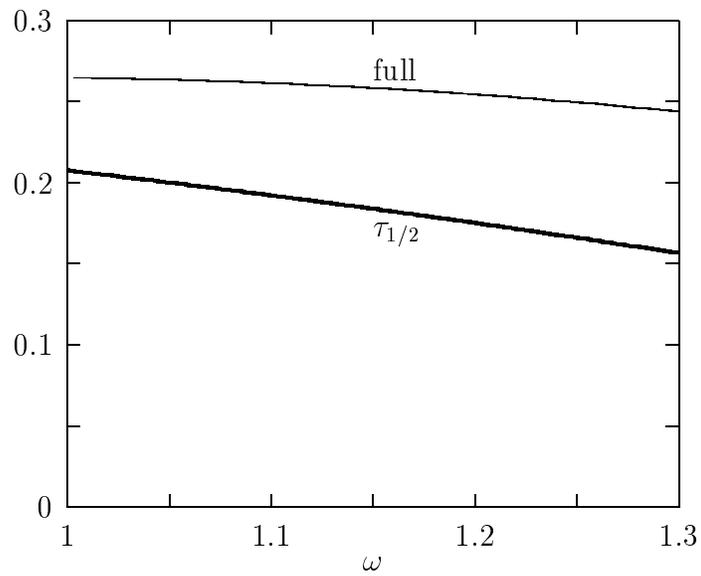
\noindent {\bf FIG. 8:}  Differential spectrum for $B\rightarrow
1\;1^+_{1/2}\ell\nu$.  Normalization is such that the spectrum becomes equal
to the Isgur-Wise function $\tau_{1/2}$ when the form factors are replaced by
their values in the heavy-quark limit. \vspace{4ex}

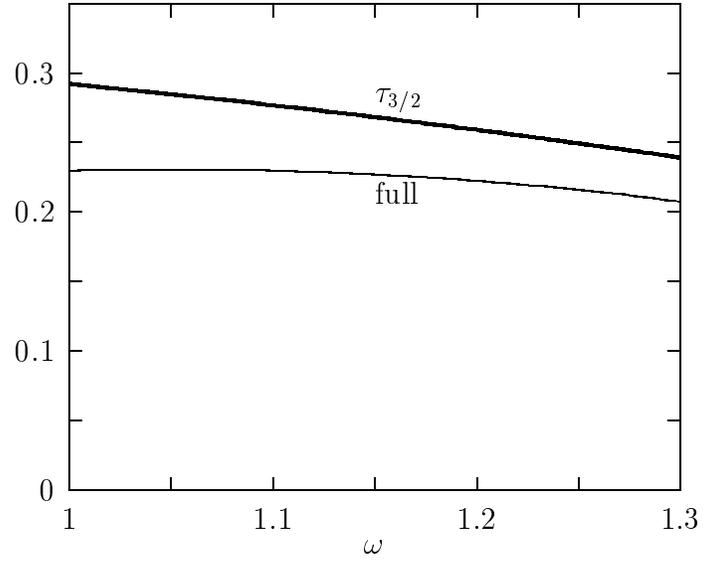
\noindent {\bf FIG. 9:} Differential spectrum for $B\rightarrow
1\;1^+_{3/2}\ell\nu$.  Normalization is such that the spectrum becomes equal
to the Isgur-Wise function $\tau_{3/2}$ when the form factors are replaced by
their values in the heavy-quark limit. \vspace{4ex}

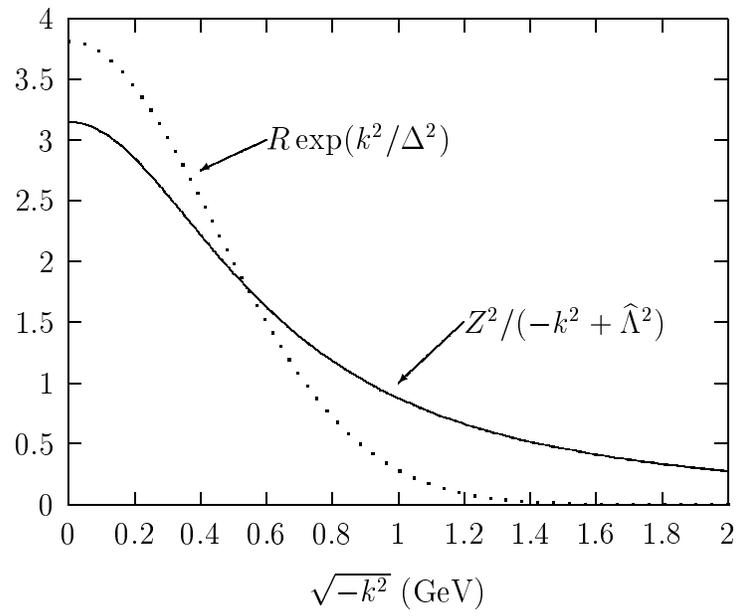
\noindent {\bf FIG. 10:} Comparison of usual damping factor with exponential
damping factor for $B$ meson, with physical values of parameters.

\end{document}